%% file: seoul.tex
\documentstyle[epsfig,rotating,seoul]{article}
\input{commandes.tex}%
\def\Journal#1#2#3#4{{#1} {\bf #2}, #3 (#4)}

\def\PLB{{\em Phys. Lett.}  B}
\def\PRL#1#2#3{{\it Phys. Rev. Lett. }{\bf #1 }(#2) #3}

\font\eightrm=cmr6

\begin{document}
\title{ 
  HIGGS SEARCHES AT LEP AND AT THE TEVATRON
  }
\author{
  Patrick Janot (E-mail: Patrick.Janot@cern.ch)\\
  {\em CERN, EP Division, CH-1211 Geneva 23} \\
  }
\maketitle
\baselineskip=11.6pt
\begin{abstract}
After years of efforts to push the LEP performance to, and indeed beyond, the 
limits of what had been believed possible, hints of a signal of a Higgs boson at
115\,\Gcs\ appeared in June 2000, were confirmed in September, and 
confirmed again in November. Spending an additional six-month period with LEP 
would have given the unambiguous opportunity of a fundamental discovery. Instead, 
this possibility was handed over to the Tevatron, for which at least six 
more years will be needed to confirm the existence of a Higgs boson around 
115\,\Gcs. The upgrades performed at LEP and needed at the Tevatron, together 
with the physics outcomes, are briefly mentioned in turn.

\end{abstract}
\baselineskip=14pt
\section{LEP (1989-2000): ``2001, A Spoilt Odyssey''}
As described in P. Renton's presentation\cite{renton}, the Luminosity, the 
Energy and the Precision (L,E,P) of the measurements made at LEP and SLD 
(available at the time of the conference) allowed an indirect prediction of 
the Higgs boson mass to be made in the framework of the standard model,
\begin{equation}
  \mH = 118^{+63}_{-42}\,\Gcs,
  \label{mH}
\end{equation}
as obtained with the as-yet most precise determination of the QED coupling 
constant evaluated at the Z mass.\cite{martin}
The prediction of such a light Higgs boson emphasized the interest of the direct 
search at LEP. 

\begin{figure}[h]
\begin{picture}(160,55)
\put(-2,7){\epsfxsize48mm\epsfbox{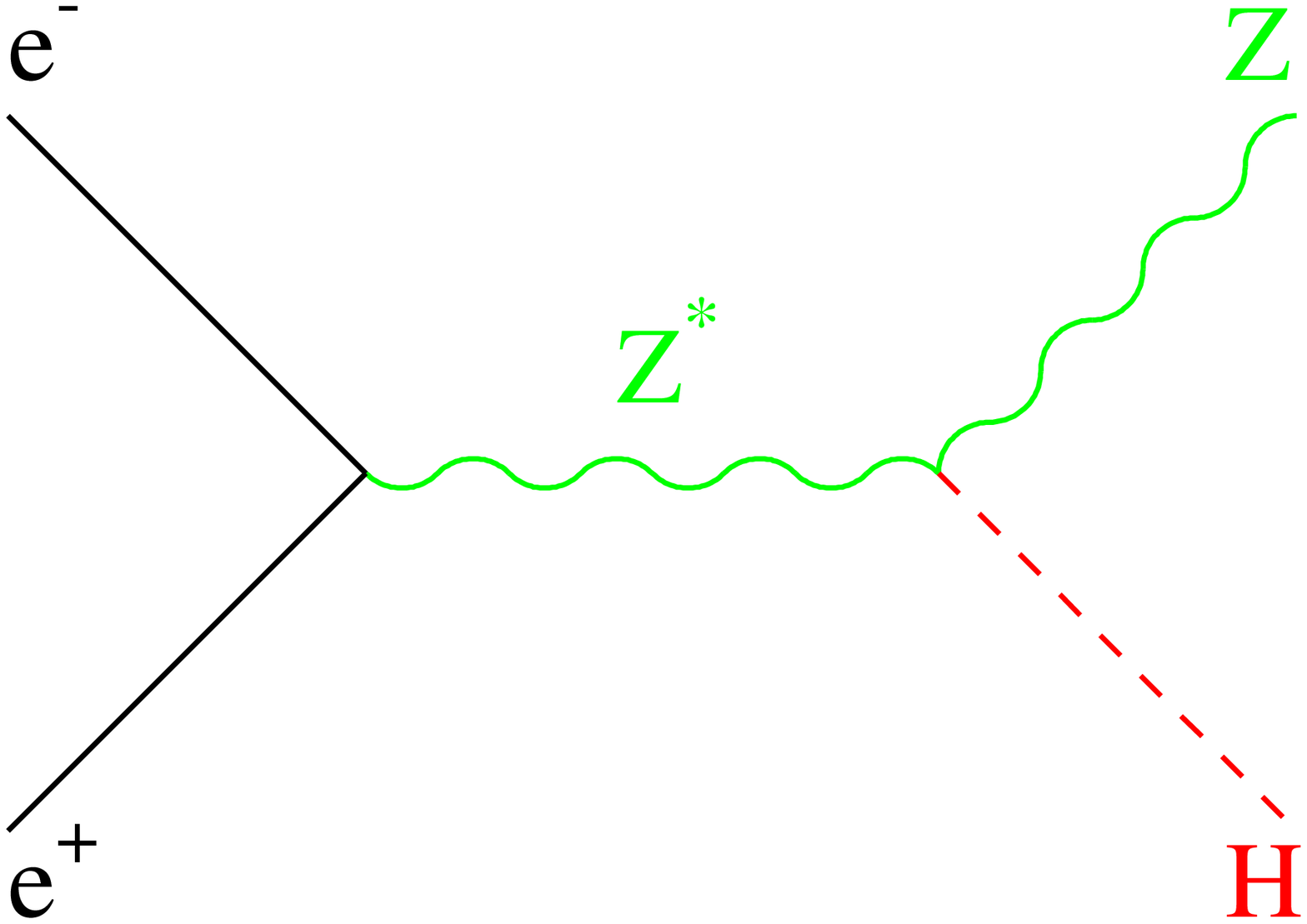}}
\put(47,-3){\epsfxsize80mm\epsfbox{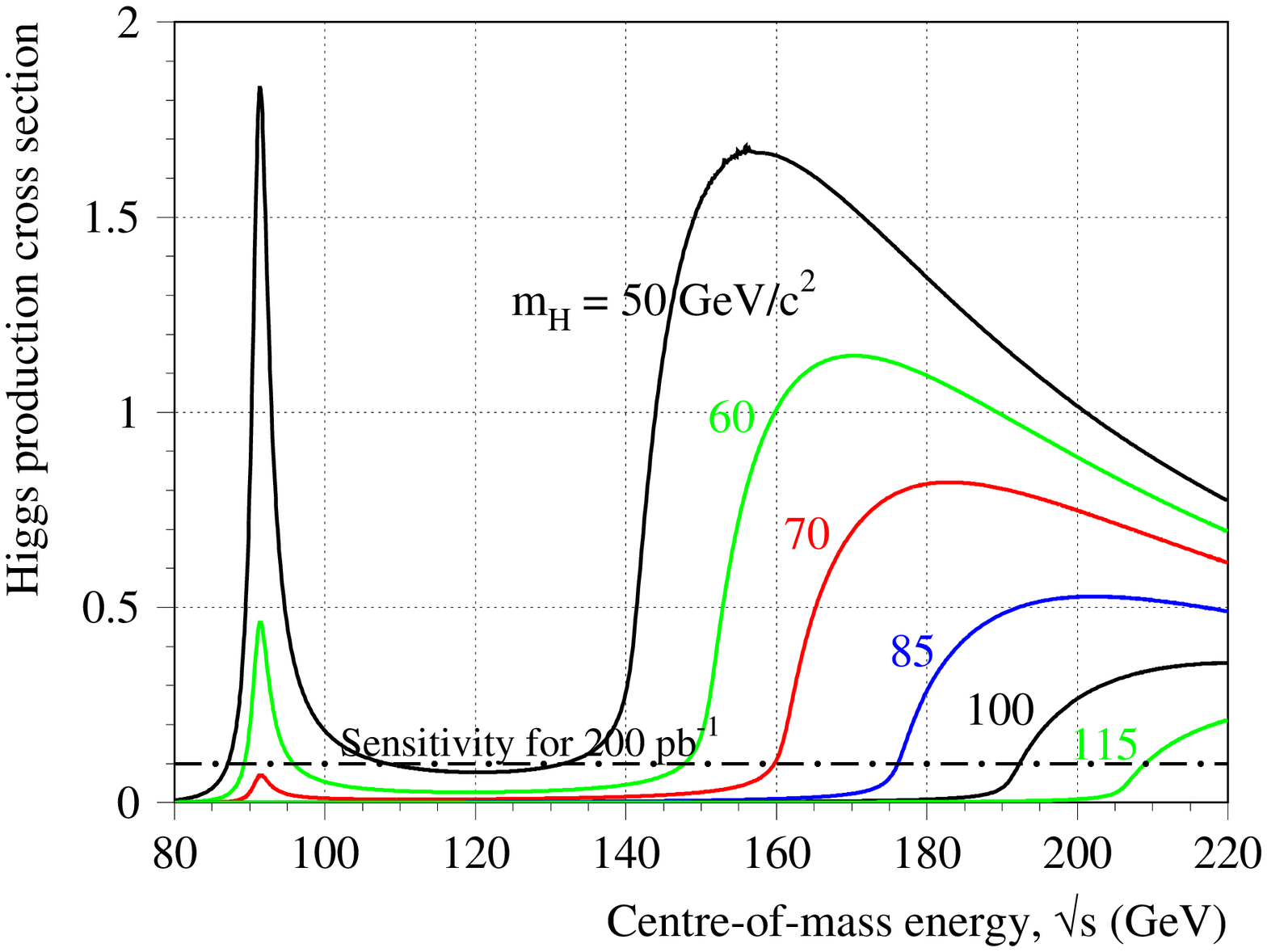}}
\end{picture}
\caption{Higgs boson production process at LEP\,2 (left) and cross section as a 
function of the centre-of-mass energy for several Higgs boson mass values. Also
indicated (dash-dotted line) is the $5\sigma$-sensitivity reached with 
200\,\inpb.}
\label{fig:crossHZ}
\end{figure}

All searches carried out during the first phase of LEP through the 
Higgstrahlung process $\epemto\ \H\ffbar$ were unsuccessful, and led to a lower 
limit of 65.6\,\Gcs\ on the standard model Higgs boson mass at 95\% C.L.\cite{gordy}
It was time, in 1995, to go to the second phase of LEP. As shown 
in Fig.\,\ref{fig:crossHZ}, a centre-of-mass energy of 192\,GeV (which was 
foreseen to be reached with the available equipment) allowed a 
$5\sigma$-sensitivity of 100\,\Gcs\ to be achieved on the standard model 
Higgs boson mass, through the search for the process $\epemto\ \H\Z$. 
Similarly, a centre-of-mass energy of 209\,GeV (actually reached in 2000) 
increased this sensitivity to 115\,\Gcs.

The search for the HZ process proceeds through three clear topologies, 
originating from the dominant decay channels of the Higgs (mostly in 
\bbbar\ for the mass range of interest at LEP) and of the Z bosons:
\begin{itemize}
\item an identified lepton pair, electrons or muons, accompanied by two b jets,
when $\Z \to \epem, \mpmm$, in less than 10\% of the cases;
\item an acoplanar pair of b jets, accompanied with missing energy and mass, 
when $\Z \to \nnbar$, in 20\% of the cases;
\item a four-jet final state when the Z decays into hadrons, in the remaining 
70\% of the existing configurations;
\end{itemize}
easily selected with efficiencies ranging from 40\% (for the four-jet final state) 
to 80\% (for the leptonic final state). However, the presence of irreducible 
backgrounds with large production cross sections (such as, e.g., $\epemto\ 
\Z\Z, \WpWm$ or $\qqbar$, which all contribute to the four-jet topology) 
requires a careful treatment on a event-by-event basis to determine the 
``signal-ness'' of each candidate. 

\begin{figure}[h]
\begin{picture}(160,90)
\put(10,-5){\epsfxsize110mm\epsfbox{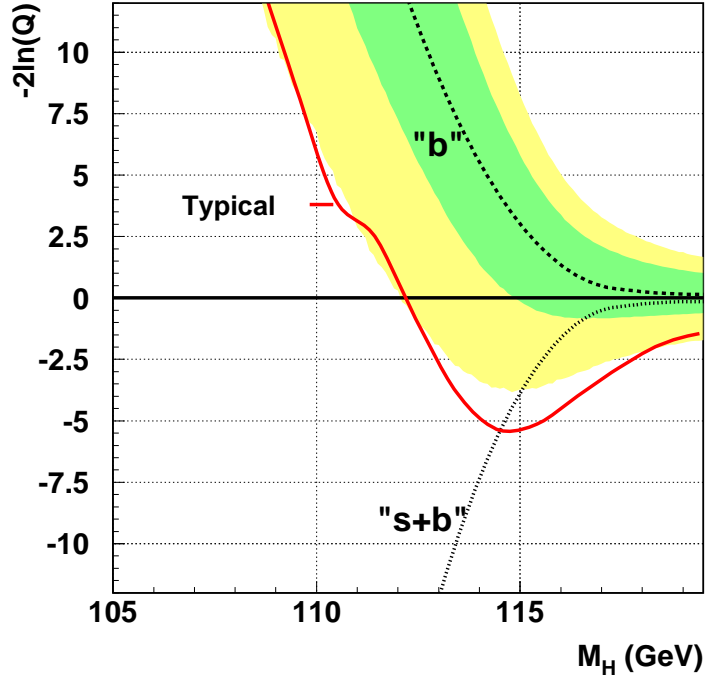}}
\end{picture}
\caption{Typical neg've log-likelihood curve of the LEP data sample for a 
Higgs boson mass of 115\,\Gcs. The dashed curves display the expected mean 
value of the minimum, should the sample contain either only background (curve 
``$b$'') or signal as well (curve ``$s+b$''), as a function of the Higgs boson mass.
The shaded bands show the 68\% and 95\% compatibility bands with the 
background-only hypothesis.}
\label{fig:like}
\end{figure}

To this end, each event was
characterized by its kinematic properties, its reconstructed mass in the 
Higgs boson hypothesis, and its b-quark content. These characteristics were 
combined with likelihood methods or neural networks, and the combined output 
was used to assign (with large simulated event samples of signal and background) 
a signal-to-noise ratio ($s/b$, Higgs-mass-hypothesis dependent) to each candidate.
The overall negative log-likelihood of a given sample of N candidate events,
\begin{equation}
L(\mH) = -2 \log Q {\rm \ \ with\ \ } Q = \prod_{i=1}^{N} \left(1 + 
{s_i \over b_i}(\mH) \right),
\label{like}
\end{equation}
smaller in presence of signal than it would be with background events 
only, was used to quantitatively estimate the result of the search. 
Because the signal cross section decreases rapidly when \mH\ increases, the 
separation between the likelihood of a signal-like and a background-like experiment
is expected to become smaller as \mH\ reaches 
the ``kinematic limit'' of HZ production, {\it i.e.}, $\mH\ \sim \sqrt{s} - \mZ$. 
The typical expected shape of $L$ as a function of the hypothetical Higgs 
boson mass, should the Higgs boson weigh 115\,\Gcs, is displayed in 
Fig.\,\ref{fig:like} for the luminosity actually recorded by the four LEP 
experiments in 2000, at centre-of-mass energies between 205 and 209\,GeV. The
minimum of the expected curve shows the most probable hypothetical mass pointed 
at by the event sample. 

With a combination of the events from the four LEP experiments, a $3\sigma$ 
sensitivity of 115\,\Gcs\ was achieved, thanks to the large integrated 
luminosity produced at and above 206\,GeV. Such high centre-of-mass energies
would not have been reached if it were not for the great ingenuity and utmost 
efforts to take advantage of all possible resources of the accelerator. Indeed, 
the existing accelerating equipment (288 Nb/Cu superconducting cavities, 
installed between 1995 and 1999, with a design accelerating gradient of 
6 MV/m) was aimed at delivering a maximum 
centre-of-mass energy of 192\,GeV. The following actions were then taken, and 
their effect on the centre-of-mass energy and the Higgs boson mass 
$3\sigma$ sensitivity are displayed in Table\,\ref{tab:effect}. 

\begin{table}[htbp]
\caption{Effect on $\sqrt{s}$ and on the $3\sigma$-sensitivity on $\mH$ of the 
various improvements brought to LEP in its last two years of running.}
\label{tab:effect}
\begin{center}
\begin{tabular}{|l|c|c|}
\hline\hline 
Action & Effect on $\sqrt{s}$ (GeV) & \mH\ sensit.(\Gcs) \\ \hline\hline
(i) Cryogenics upgrade & $192 \to 204$ &  $100 \to 112$ \\ \hline 
(ii) One klystron margin & $204 \to 205.5$ &  $112 \to 113$ \\ \hline 
(iii) Mini-ramps to no margin & $205.5 \to 207$ &  $113 \to 114$ \\ \hline 
(iv) Eight Cu cavities & $207 \to 207.4$ &  $114 \to 114.25$ \\ \hline 
(v) Orbit correctors & $207.4 \to 207.8$ &  $114.25 \to 114.5$ \\ \hline 
(vi) Smaller RF frequency & $207.8 \to 209.2$ &  $114.5 \to 115.1$ \\ \hline\hline
\end{tabular}
\end{center}
\end{table}

\begin{enumerate}
\item The cryogenic installation was upgraded (as foreseen for the LHC)
to allow the accelerating gradient of the superconducting cavities to be 
gradually increased from 6 MV/m to 7.5 MV/m, for a global gain of 650 MV. 
The overall stability of the cryogenic system was also greatly improved with 
this upgrade.
\item With this gain in stability, the RF margin was reduced from 200 MV 
(corresponding to a margin of two klystrons allowed to trip without losing 
the beams) to 100 MV (only one klystron margin) with only moderate a 
reduction of the average fill duration.
\item At the end of each fill, mini-ramps to a no-margin situation were
performed, allowing another 100 MV to be gained for a duration of 
approximately fifteen minutes (the average time between two klystron trips).
\item Eight warm Cu cavities (from the first phase of LEP) were re-installed
for an additional of  gain of 30\,MV.
\item Unused (mostly uncabled) orbit correctors were powered in series to act 
as magnetic dipoles, thus increasing the bending length of LEP and allowing
the beam energy to be increased while keeping constant the energy loss by 
synchrotron radiation .
\item The radio-frequency was slightly reduced (by 100 Hz out of 350 MHz), to 
benefit from the dipolar magnetic field seen by the beam in the focusing 
quadrupoles and from the additional margin brought by the resulting shortening 
of the bunches.
\end{enumerate}

\begin{figure}[htbp]
\begin{picture}(160,95)
\put(5,-32){\epsfxsize110mm\epsfbox{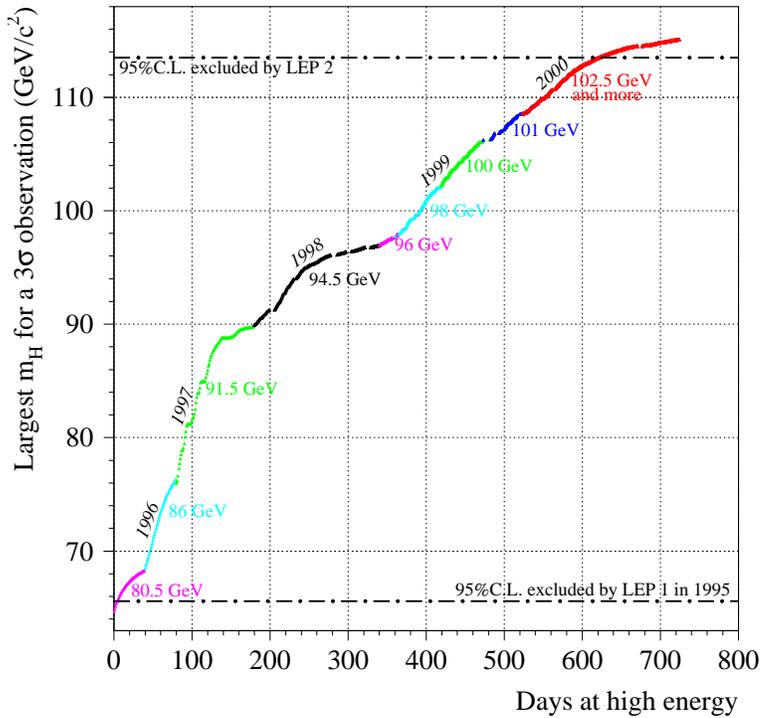}}
\end{picture}
\caption{Evolution of the $3\sigma$-sensitivity on \mH (and of $\sqrt{s}$) 
from 1996 to 2000.}
\label{fig:increase}
\end{figure}

\noindent
Altogether, these improvements allowed the maximum centre-of mass energy to be 
raised from 192 to 209.2 GeV, and the $3\sigma$-sensitivity on the standard model 
Higgs boson mass to be increased from 100 to 115.1\,\Gcs. The evolution of the
sensitivity as a function of time since 1996, displayed in Fig.~\ref{fig:increase}, 
is essentially driven by the number of superconducting cavities installed in LEP
(176 in 1996, 240 in 1997, 272 in 1998 and 288 in 1999). It is worth 
noting that 372 cavities ({\it i.e.}, as many as could possibly be installed in 
the LEP tunnel) would have allowed a large integrated luminosity to be 
produced at centre-of-mass energies in excess of 220\,GeV.

Because, until June 2000, no noticeable excess of signal-like candidate events
had been seen in the LEP data, the whole \mH\ range between 0 and 114.1\,\Gcs\
was excluded at the 95\% confidence level. In June 2000, sizeable luminosity
at centre-of-mass energies above 206\,GeV ({\it i.e.}, above the kinematic threshold 
for a Higgs boson of 115\,\Gcs) started to be steadily delivered. From 
this moment onwards, signal-like events compatible with the production of a 
Higgs boson with mass 115\,\Gcs\ were regularly recorded by the LEP experiments.
The reconstructed masses of the fourteen most significant events (selected 
with a cut corresponding to an integrated signal-to-noise ratio of about 1.0), 
their $s/b$ values, the topologies and the experiments in which they were 
detected are summarized in Table~\ref{tab:events}.

\begin{table}[htbp]
\caption{Signal-to-noise $s/b$ at 115\,\Gcs, reconstructed Higgs mass (in \Gcs), 
final state channel and experiment for the fourteen most signal-like events (selected 
with $s/b > 0.3$) corresponding to an expected purity of 50\%.}
\label{tab:events}
\begin{center}
\begin{tabular}{|c|c|c|c|}
\hline\hline 
$s/b$ & Rec. mass & Channel & Experiment \\ \hline\hline
4.7   & 114 & H\qqbar & ALEPH \\ \hline 
2.3   & 112 & H\qqbar & ALEPH \\ \hline 
2.0   & 114 & H\nnbar & L3 \\ \hline 
0.90  & 110 & H\qqbar & ALEPH \\ \hline 
0.60  & 118 & H\epem & ALEPH \\ \hline 
0.52  & 113 & H\qqbar & OPAL \\ \hline 
0.50  & 111 & H\qqbar & OPAL \\ \hline 
0.50  & 115 & H\tptm & ALEPH \\ \hline 
0.50  & 115 & H\qqbar & ALEPH \\ \hline 
0.49  & 114 & H\nnbar & L3 \\ \hline 
0.47  & 115 & H\qqbar & L3 \\ \hline 
0.45  &  97 & H\qqbar & DELPHI \\ \hline 
0.40  & 114 & H\qqbar & DELPHI \\ \hline 
0.32  & 104 & H\nnbar & OPAL \\ \hline\hline
\end{tabular}
\end{center}
\end{table}

The characteristics of the fourteen events displayed in Table~\ref{tab:events} are 
those determined as of November 2000. (A more recent update was not available at 
the time of the conference; no final update exists either at the time of writing.) 
Seven background events were expected in this data sample (and therefore seven 
signal events, should the Higgs boson weigh 115\,\Gcs), in close agreement with 
the number of events observed. It is important to note that this agreement is 
independent of the $s/b$ cut chosen, {\it i.e.}, on the expected signal purity of the event 
sample. In addition, the fourteen events are divided into
\begin{itemize}
\item Nine four-jet (H\qqbar) candidate events (expected fraction 70\%);
\item Three missing energy (H\nnbar) candidate events (expected fraction 20\%);
\item Two leptonic (H\lplm, H\tptm) candidate events (expected fraction
10\%);
\end{itemize}
in close agreement with the expected HZ fractions, and into
\begin{itemize}
\item Six events in ALEPH;
\item Three events in OPAL;
\item Three events in L3;
\item Two events in DELPHI;
\end{itemize}
to be compared with $\sim 1.7$ background events expected in each experiment. 
Such distribution is, for these small statistics, well compatible with a democratic 
production in the four LEP experiments.

\begin{figure}[htbp]
\begin{picture}(160,150)
\put(7,56){\epsfxsize100mm\epsfbox{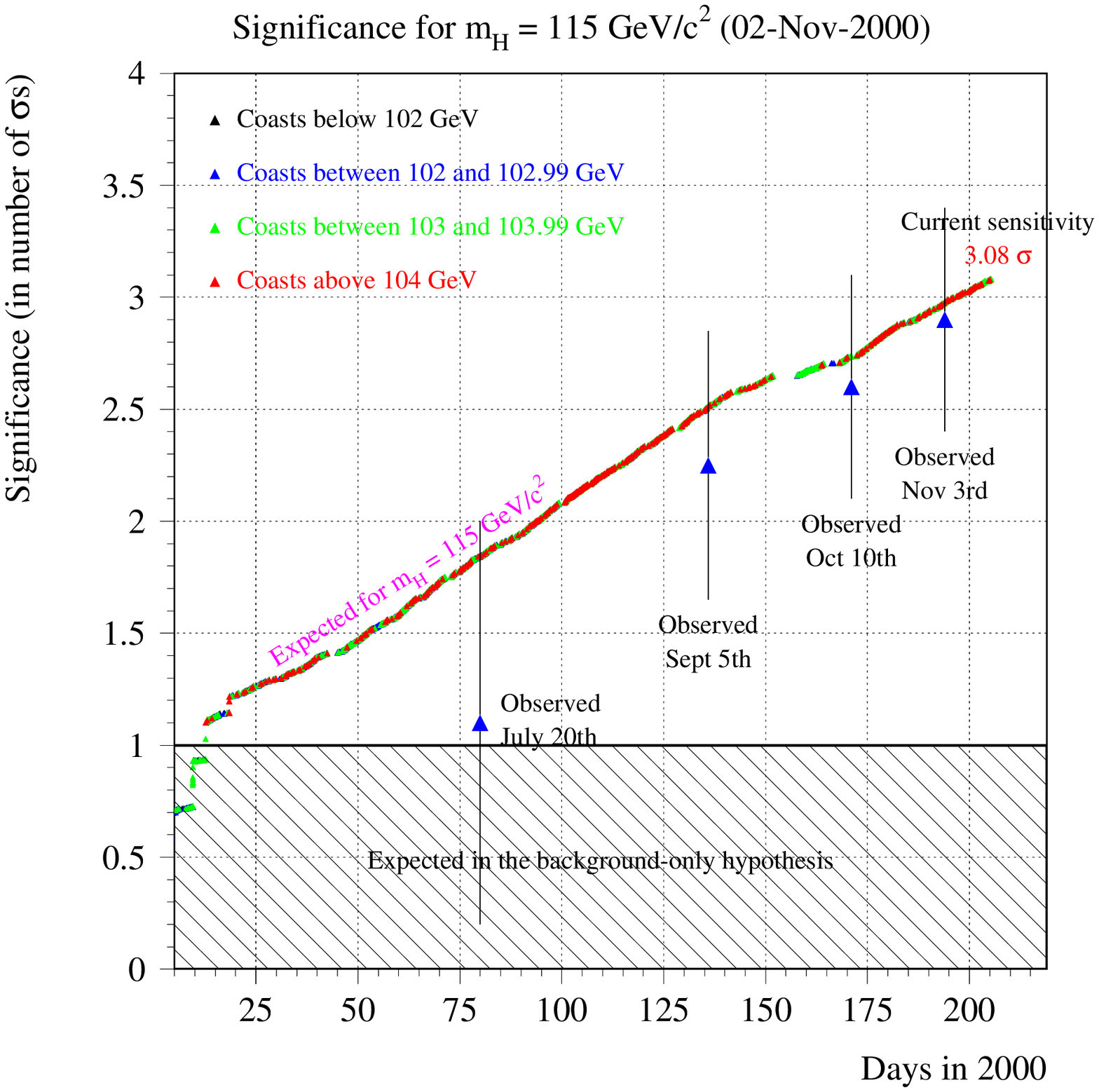}}
\put(5,-6){\epsfxsize100mm\epsfbox{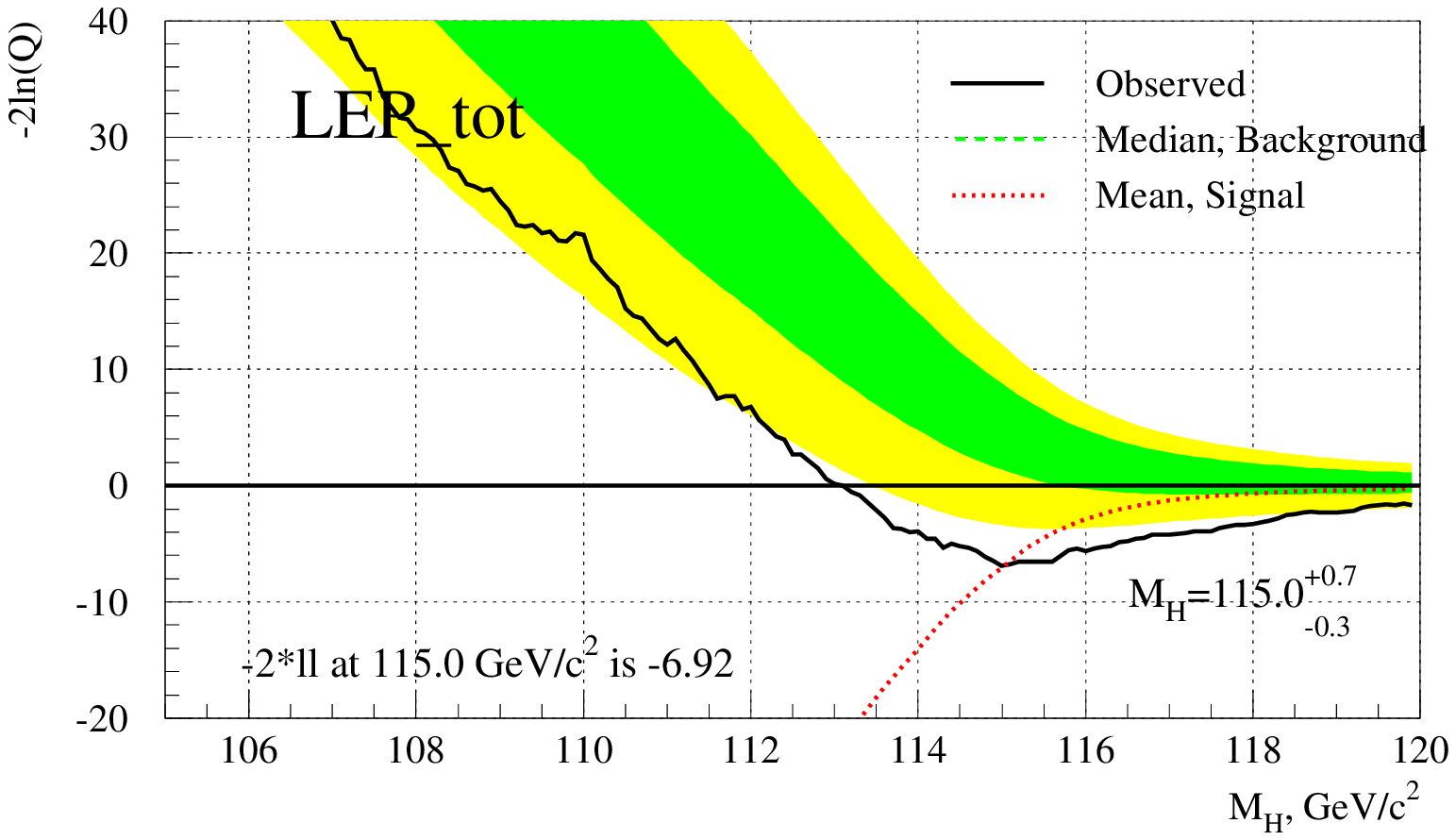}}
\end{picture}
\caption{\footnotesize Top: Increase of the observed combined significance at 
$\mH = 115\,\Gcs$ in 2000, compared with an online estimate of the  
significance expected in the signal-plus-background hypothesis. 
Bottom: Negative log-likelihood as a function of 
the hypothetical standard model Higgs boson mass. 
\label{fig:final}} 
\end{figure}

The overall observation therefore shows an impressive consistency with the signal 
hypothesis with $\mH = 115\,\Gcs$, regarding the total cross-section, 
the distribution in the four experiments and in the three final states, 
and the distribution of $s/b$. The increase of the excess significance closely 
followed, since June 2000, that expected from the presence of a 115\,\Gcs\ 
Higgs boson, as shown in Fig.~\ref{fig:final}. The final negative log-likelihood, 
with a minimum, corresponding to 2.9 standard deviations away from the background
expectation, at 
\begin{equation}
\label{mHdirect}
\mH\ = 115^{+0.7}_{-0.3}\,\Gcs,
\end{equation}
is also displayed in Fig.~\ref{fig:final}. More details, figures and cross-checks, 
further showing the robustness of the interpretation, are discussed in 
Refs.\cite{lephwg,qanda}. 

In a preliminary update released after the conference,\cite{prelim} with data 
reprocessed from L3, ALEPH and OPAL, and with additional systematic studies, 
the excess of signal-like events is still present, at a mass of 115.6\,\Gcs. 
(A 2.9$\sigma$ excess is still observed by those three experiments, slightly damped 
by DELPHI's unreprocessed, preliminary data.) In particular, the presence of 
the events with the largest $s/b$ values is confirmed. Unfortunately, if the 
Higgs boson weighs 115\,\Gcs, it is not before an \epem\ linear collider starts 
producing HZ data that events with such a high purity will be seen again. 

\begin{figure}[htbp]
\begin{picture}(160,55)
\put(-5,-2){\epsfxsize70mm\epsfbox{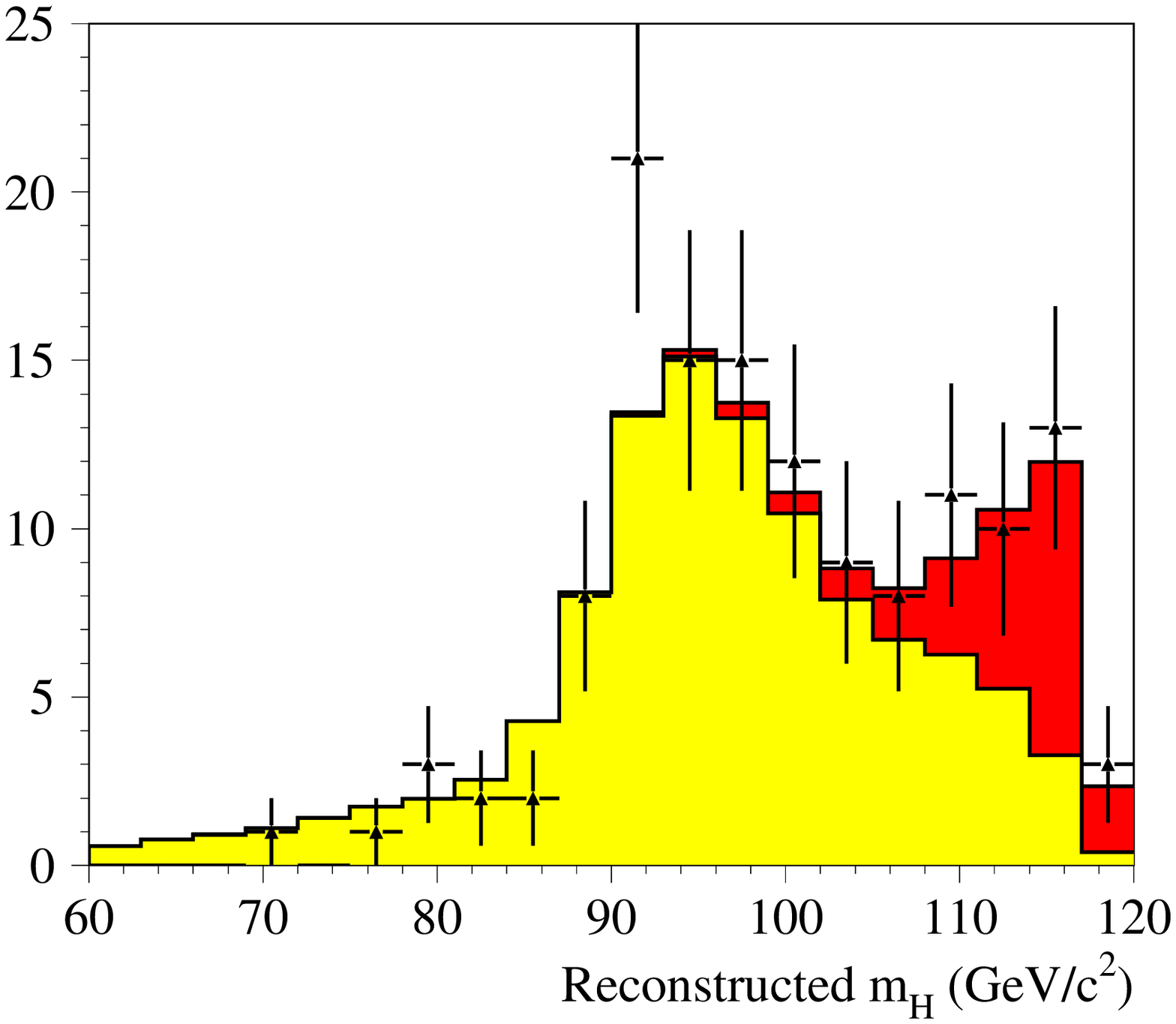}}
\put(57,-2){\epsfxsize70mm\epsfbox{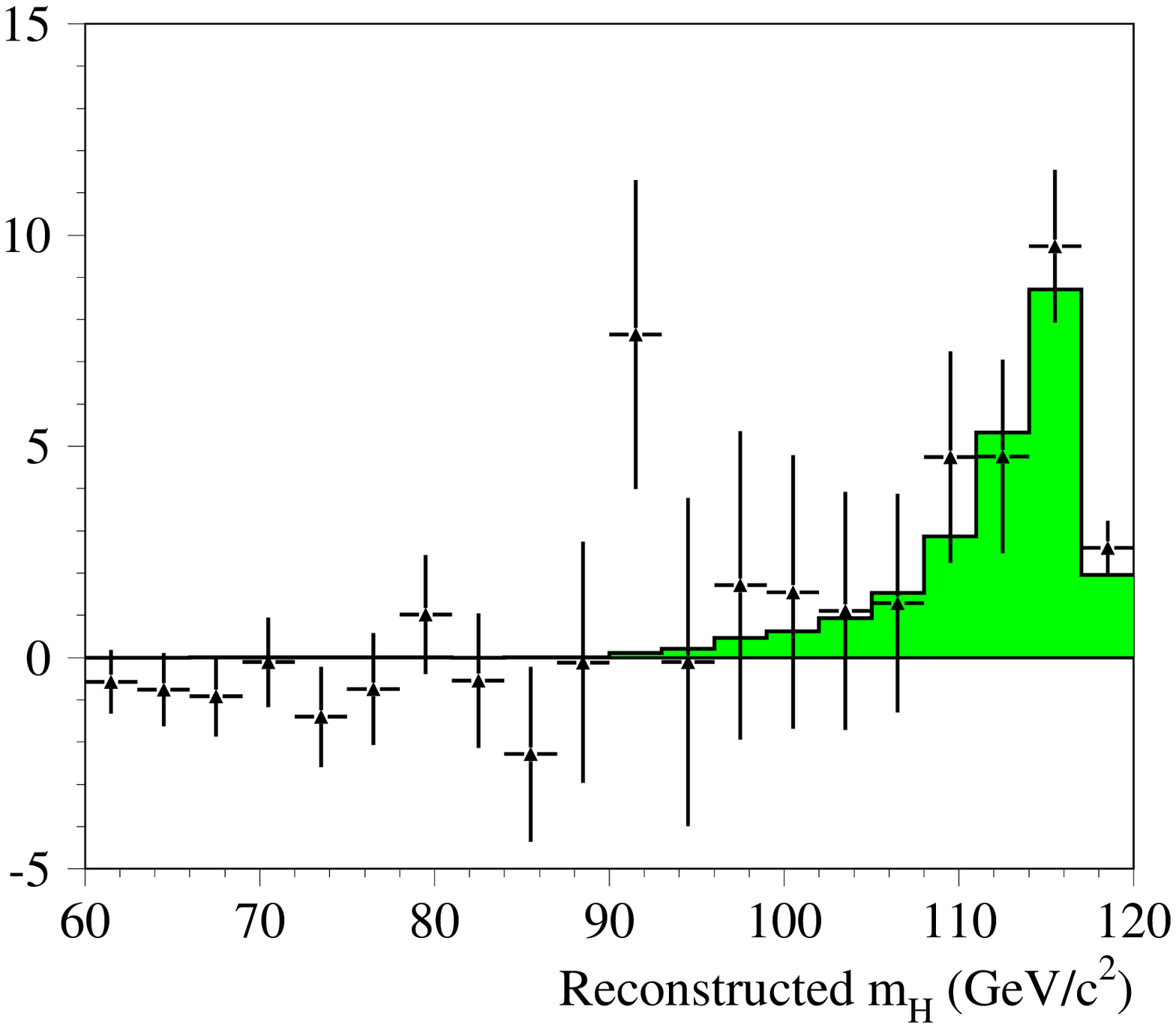}}
\end{picture}
\caption{\footnotesize Expected reconstructed mass spectrum of the most significant 
events (with an $s/b$ value in excess of 0.5) after a six-month run of LEP in 2001, 
should the Higgs boson mass indeed be around 115\,\Gcs. Left: Raw spectrum; Right:
Background subtracted spectrum, with an expected excess of 28 events.
\label{fig:recmh}} 
\end{figure}

With six more  months of LEP running in 2001, {\it i.e.}, with an integrated luminosity 
of 200\,\inpb\ and an upgraded centre-of-mass energy above 208.5 GeV (made possible 
with a few available additional cavities and few accelerator tricks), the almost 
3$\sigma$ excess could have turned into an unambiguous $5.5^{+0.6}_{-0.9} \sigma$ 
discovery, and have led to the reconstructed mass spectrum displayed in 
Fig.~\ref{fig:recmh}, should the Higgs boson mass indeed be around 115\,\Gcs. 
Similarly, in the null hypothesis, the new data would have allowed to 
demonstrate that the excess seen in 2000 was due to a statistical fluctuation.

\noindent 
In contrast, CERN's Director General decided to shut down LEP for ever on 
November 17th, 2001, at 4:15pm.

\section{Tevatron (2001-2007++): ``Back to the future''}
Next-in-line for the Higgs boson search is the Tevatron. Run\,2 started nearly at the 
time of the conference, and is supposed to last at least until LHC starts delivering
useful data for this search ({\it i.e.}, at least until 2007). About 0.1\,\infb\ of data 
were collected by both experiments, CDF and D0, during the Run\,1 at a 
centre-of-mass energy of 1.8\,TeV. The goal of Run\,2 is to increase this 
figure to 2, 4 and 15\,\infb\ in 2003, 2004 and 2007 respectively, at an 
upgraded 2 TeV centre-of-mass energy.

\begin{figure}[htbp]
\begin{picture}(160,78)
\put(15,40){\epsfxsize38mm\epsfbox{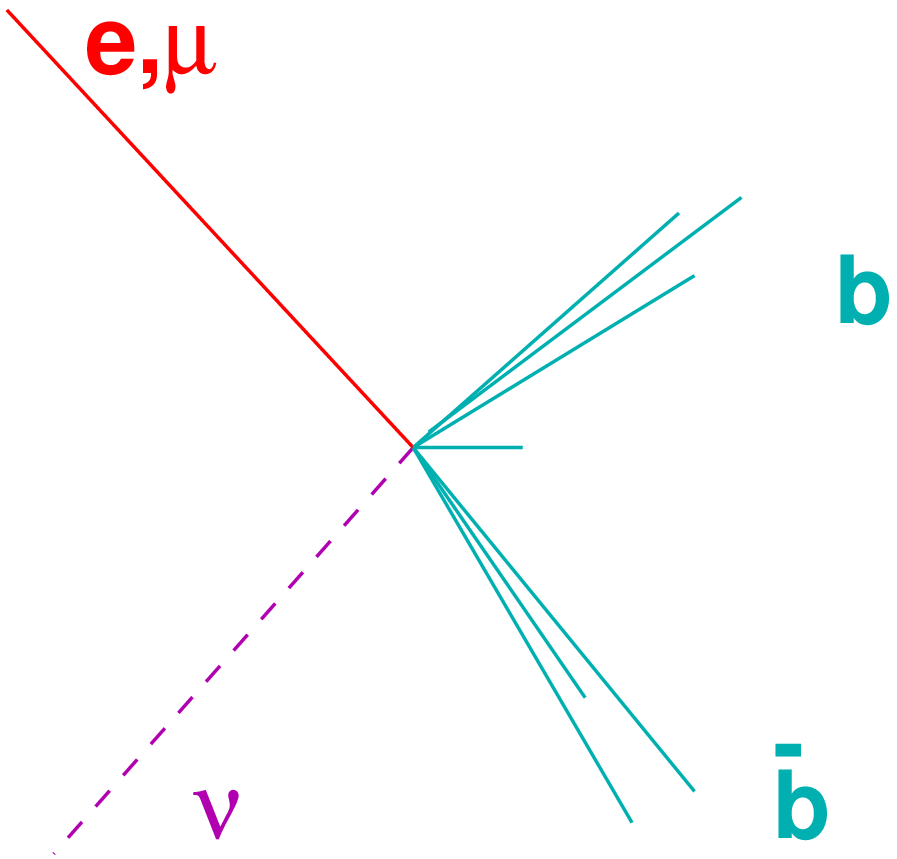}}
\put(70,40){\epsfxsize43mm\epsfbox{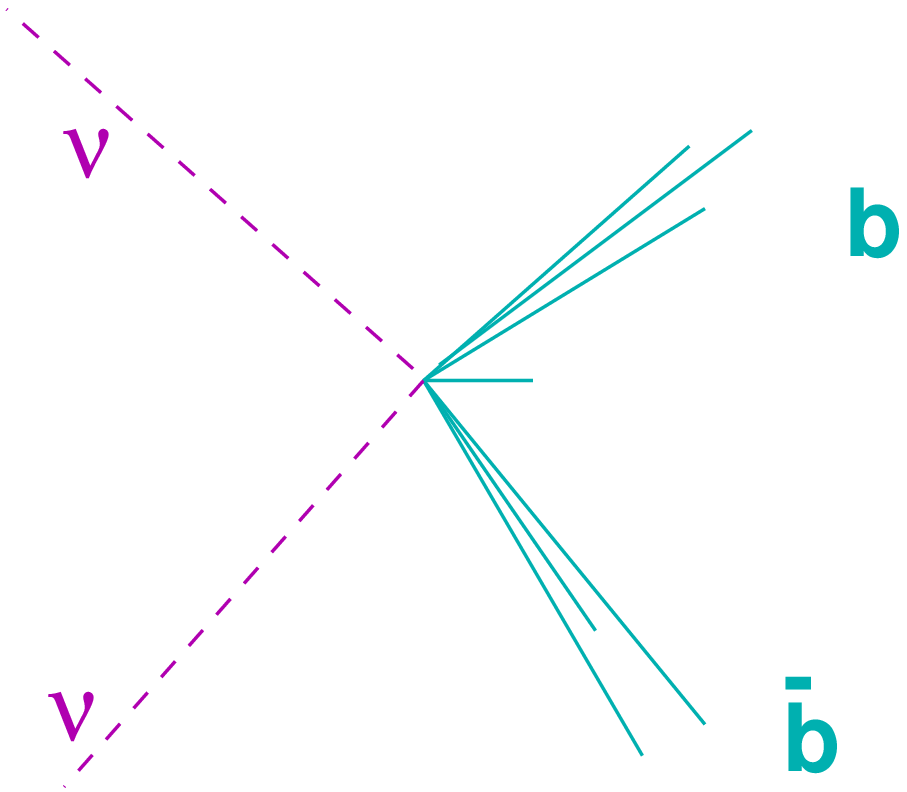}}
\put(15,0){\epsfxsize38mm\epsfbox{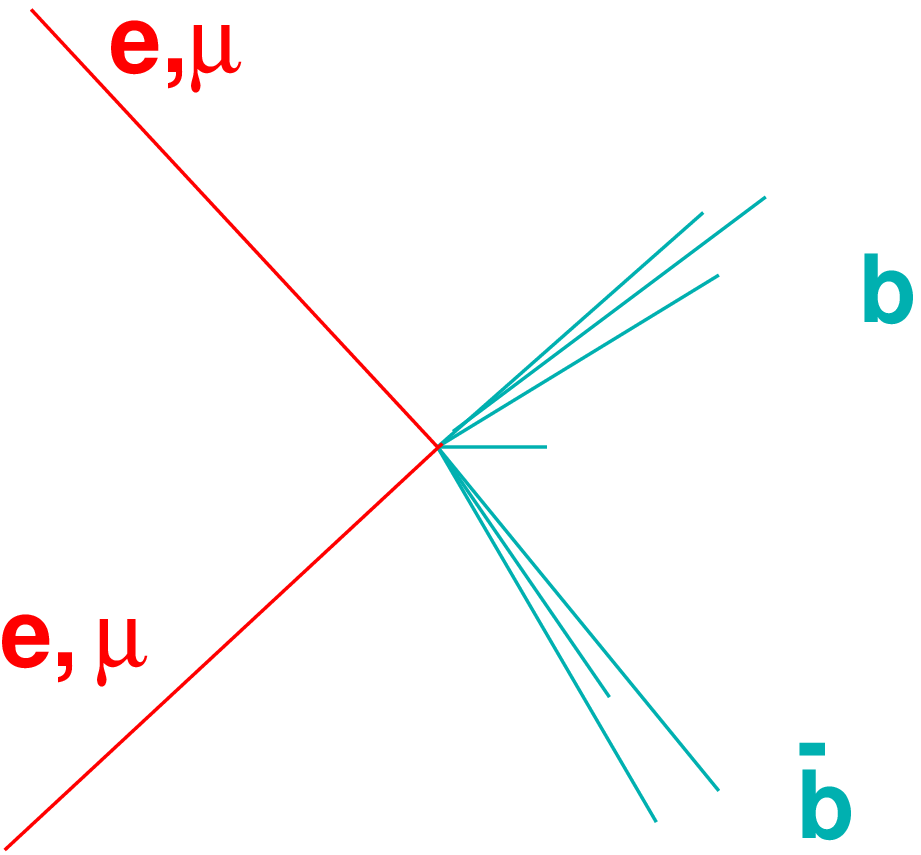}}
\put(70,0){\epsfxsize48mm\epsfbox{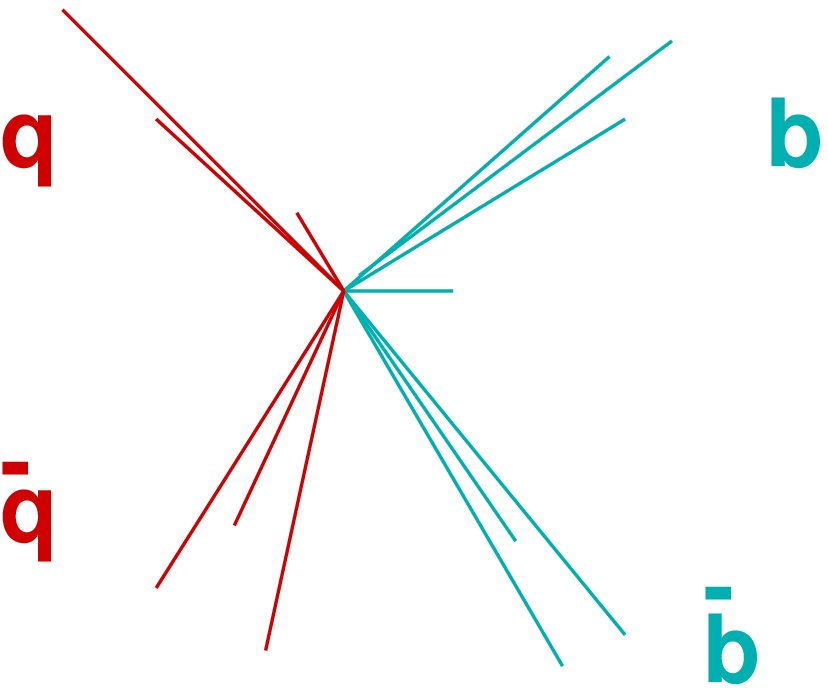}}
\end{picture}
\caption{Final states for $\mH < 130\,\Gcs$, with $\H\to\bbbar$. Other final states
are studied for $\mH > 130\,\Gcs$, with $\H\to\W\W^\ast$.}
\label{fig:hsmtevb}
\end{figure}

The dominant Higgs production process in \ppbar\ collisions at $\sqrt{s} = 2$\,TeV 
is the gluon-gluon fusion $\g\g\to\H\ \to\bbbar$. However, due to the overwhelming 
dijet QCD background, the only practicable processes are similar to that dominant 
at LEP, $\qqbar \to \H\Z$ and $\qqpbar \to \H\W$. Beyond LEP sensitivity, the 
cross sections of these processes are below 0.2\,pb. The possible final states, 
for a Higgs boson mass below 130\,\Gcs\ (for which $\H\ \to \bbbar$ dominates) 
are sketched in Fig.~\ref{fig:hsmtevb}. 

\noindent

\begin{figure}[t]
\begin{picture}(160,91)
\put(12,-15){\epsfxsize95mm\epsfbox{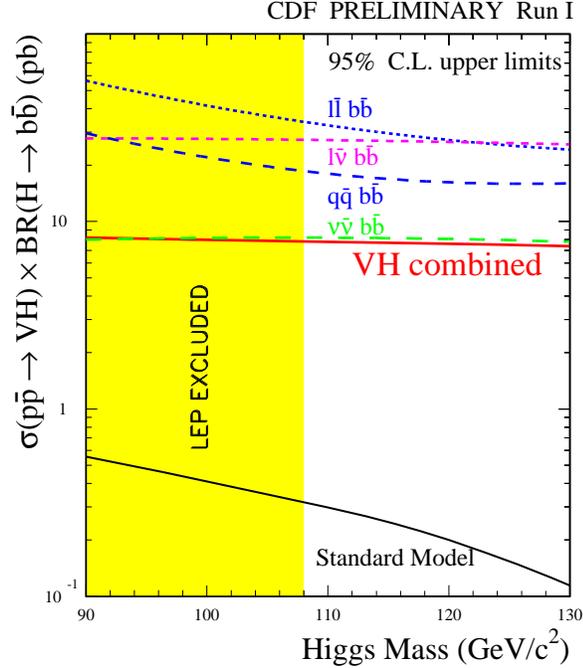}}
\end{picture}
\caption{Upper limit set by CDF on the HV production cross sections
times the $\H\to\bbbar$ branching fraction with data taken in 
Run\,1, compared with the standard model prediction. The comparison with 
LEP limits is incorrectly quoted here and in Ref.\cite{cdfhsm}. A proper 
comparison as a function of $\sigma\times BR$ is shown in Ref.\cite{transp}.}.
\label{fig:cdflimit}
\end{figure}

These four final states have already 
been studied at the Tevatron in Run\,1. The CDF reconstructed Higgs boson mass 
distributions\cite{cdfhsm} show no apparent excess over the expected background. 
However, the sensitivity is well short of the standard model expectations: 
a 95\% C.L. upper limit of about 7\,pb was set on the HV (where V stands 
for W and Z) production cross section times the $\H\to\bbbar$ branching 
fraction, to be compared with a standard model cross section  of 0.25\,pb for 
$\mH = 115\,\Gcs$ (see Fig.~\ref{fig:cdflimit}).

The missing factor of 27 in sensitivity corresponds to a factor of 700 
in effective integrated luminosity, {\it i.e.}, CDF alone would need 
$\sim 70\,\infb$ (resp. 450\,\infb) to achieve a 95\% C.L. (resp. 5$\sigma$)
sensitivity for $\mH = 115\,\Gcs$  if nothing had been changed either to the 
detector or to the analyses. The upgrades envisioned to reduce the needs down 
to a couple of \infb\ (resp. 15\,\infb) are listed in Table~\ref{tab:RunII} 
and briefly addressed in turn in the following.

\begin{table}[htbp]
\caption{Upgrades envisioned to reduce the needs in integrated luminosity 
for the search for the standard model Higgs boson in Run\,2. The expected 95\% C.L. 
limit on the HV production cross section, the integrated luminosity needed per 
experiment to exclude a 115\,\Gcs\ Higgs boson, and that needed to find it are 
given.}
\label{tab:RunII}
\begin{center}
\begin{tabular}{|l|c|c|c|}
\hline\hline 
Improvement & $\sigma_{95}$ (pb) & ${\cal L}_{95}$ (\infb) 
& ${\cal L}_{5\sigma}$ (\infb)\\ \hline\hline
Now (CDF, 0.1 \infb)          & 6.0 & 70  & 450 \\ \hline 
2 expts ($\times 2$)          & 4.2 & 35  &  220 \\ \hline 
$\sqrt{s} = 2$\,TeV $(+30\%)$ & 3.7 & 25  & 150 \\ \hline 
Lepton acc. $(+30\%)$         & 3.3 & 20  & 125 \\ \hline 
b tagging eff. $(+50\%)$      & 2.7 & 15  &  95 \\ \hline 
Mass resolution $(-30\%)$     & 2.0 &  8  &  50 \\ \hline 
NN analyses eff. $(+30\%)$    & 1.6 &  5  &  35 \\ \hline 
Trigger eff. ($\times 2$)     & 1.1 & 2.5 &  15 \\ \hline 
\end{tabular}
\end{center}
\end{table}

\subsection{\it Two experiments for Higgs search?}
\label{sec:twoexpt}

Most of the D0 subdetectors are new with respect to Run\,1. In particular, a 
superconducting solenoid and a silicon microstrip tracker were installed which 
will greatly improve the b-tagging capabilities (and therefore the Higgs boson 
search efficiency) of D0. Many other new components will allow D0 to catch up on 
and possibly exceed CDF performance.

\subsection{\it Tevatron energy upgrade}
\label{sec:energy}

The Run\,2 beam energy goal is 980\,GeV ($\sqrt{s} = 1.96$\,TeV). A heavy programme
of cryogenics upgrade (with central helium liquefier upgrades, installation of
more cold compressors and heat exchangers, swapping of the weakest magnets to 
the coldest regions, \dots.) over the past eight years made this upgrade possible. 
The Tevatron ran successfully at 980\,GeV in the $1\times8$ bunch 
configuration on April 3-5, and the design $36\times36$ bunch configuration. All 
dipoles were ramped up to 1010\,GeV, and all low-$\beta$ quadrupoles to 1030\,GeV, 
thus leaving a comfortable margin for operations at 1.96\,TeV.

\subsection{\it Lepton Id and b-tagging acceptance}
\label{sec:lepbtg}

The CDF detector also underwent major upgrades in the past five years, with a brand 
new eight layers silicon tracking system (of which three layers down to a 
pseudo-rapidity of 3.0), new end-plug calorimeters and forward muon detectors, 
extending full electron and muon coverages down to $\vert \eta \vert = 3.6$ and
1.5, respectively. The b-tagging and lepton-Id coverage is similar in D0.

\subsection{\it Other detector and analysis improvements}
\label{sec:other}

Other potential improvements are worth a factor of 6 in 
integrated luminosity, but they remain to be carefully worked out. First, the dijet
mass resolution ought to be improved from 15 to 10\%, which requires good and
constant energy-flow capabilities. Algorithms are currently being thought of 
and developed. Second, analysis efficiencies are hoped to be increased by 30\% by 
neural network techniques and by designing more subtle selections than those 
described in Ref.\cite{RunII}. Finally, only future will tell if trigger 
efficiencies (especially b-jet triggers, particularly difficult to simulate) 
can indeed be doubled.

\subsection{\it Results and Luminosity needed}
\label{sec:result}

\begin{figure}[htbp]
\begin{picture}(160,150)
\put(10,59){\epsfxsize98mm\epsfbox{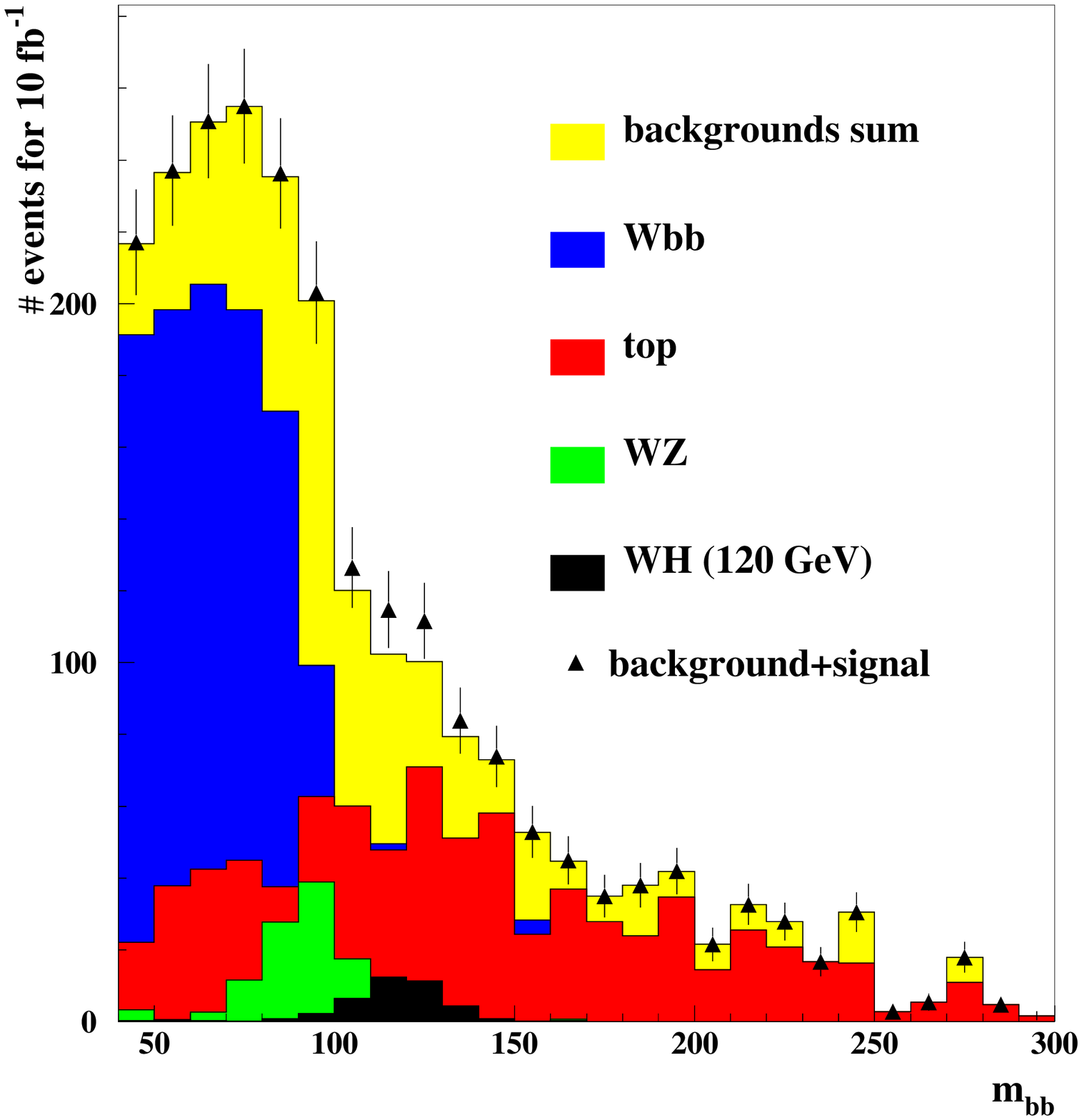}}
\put(8,-2){\epsfxsize100mm\epsfbox{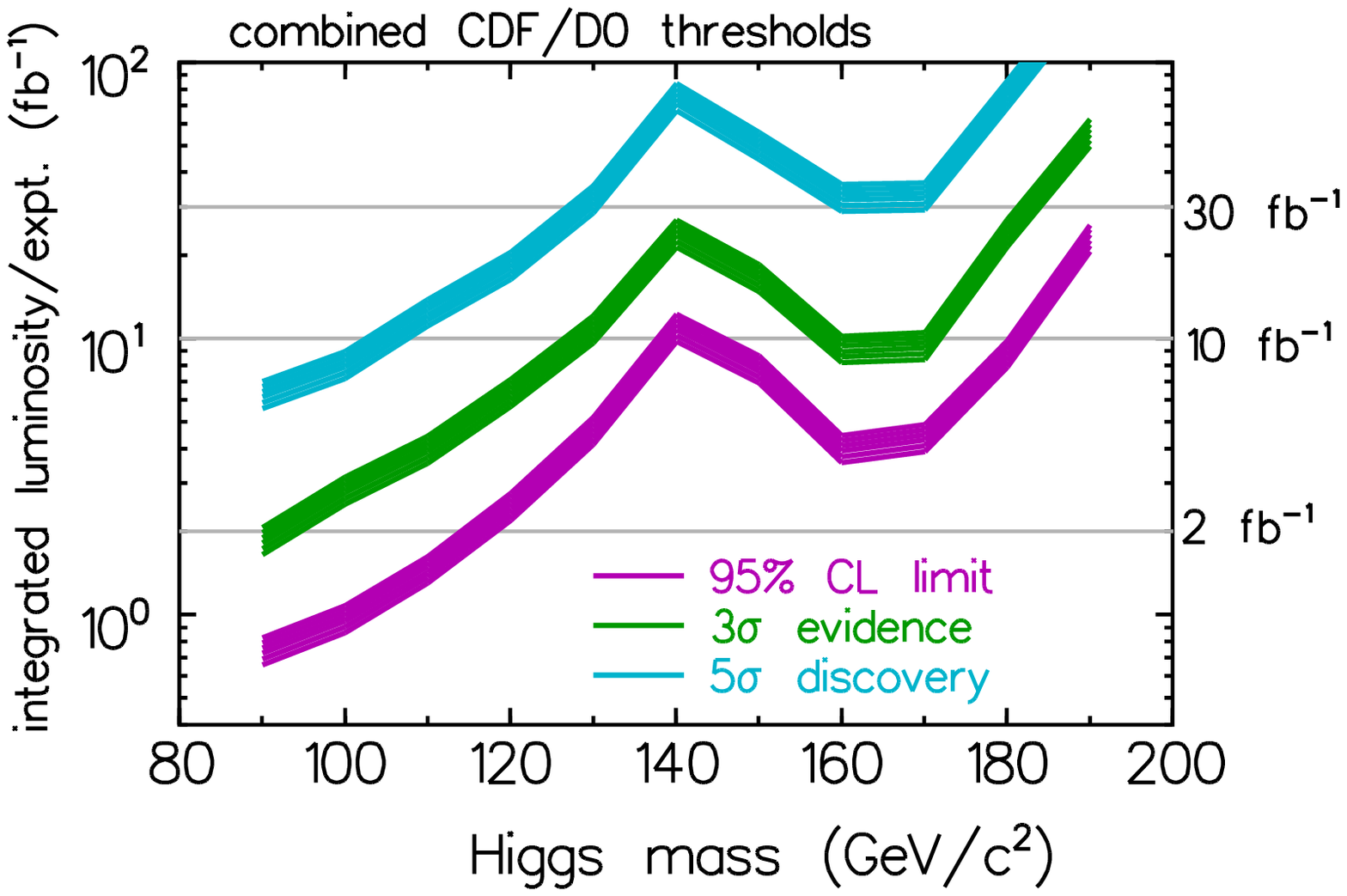}}
\end{picture}
\caption{\footnotesize Top:  \bbbar\ mass distribution in the 
$\H\W \to \bbbar\ell\nu$ final state, with 10\,\infb\ and $\mH = 120\,\Gcs$.
The histograms represent the contributions of the background processes.
The triangles with error bars include that of the Higgs boson. 
Bottom: Minimum luminosity needed to exclude
(bottom curve) or discover (top curve) a standard model Higgs boson;
\label{fig:conway}} 
\end{figure}

Taking into account all the above improvements, the expected \bbbar\ mass 
distribution for the most copious channel ($\W\H\ \to \ell\nu\bbbar$) 
with 10\,\infb\ and for $\mH\ = 120\,\Gcs$, is shown in Fig.~\ref{fig:conway}. 
Once combined with all other channels, 
the luminosity needed to reach a 95\% C.L. exclusion, 3$\sigma$ observation
or 5$\sigma$ discovery sensitivity is displayed in the same figure as a function
of the hypothetical Higgs boson mass. 

For a Higgs boson mass of 115\,\Gcs, a significance of 2$\sigma$ is expected to be 
reached in 2003 with 2.5\,\infb, 3$\sigma$ in 2005 with 5\,\infb
and 5$\sigma$ in 2007 with 15\,\infb.\cite{RunII}  These figures were revisited  
by independent LHC studies at 2\,TeV\cite{fab}, and were found to be slightly 
smaller (1, 2 and 3$\sigma$, respectively). While a 3$\sigma$ hint would certainly 
be enough to convince the community of the existence of a 115\,\Gcs\ Higgs boson 
(it would then be a confirmation of LEP's hints), the situation becomes much more 
difficult above 115-116\,\Gcs\ for which a 5$\sigma$ signal would be needed to 
claim a discovery.

\subsection{\it Luminosity upgrades}

\begin{figure}[h]
\begin{picture}(160,85)
\put(0,0){\epsfysize85mm\epsfbox{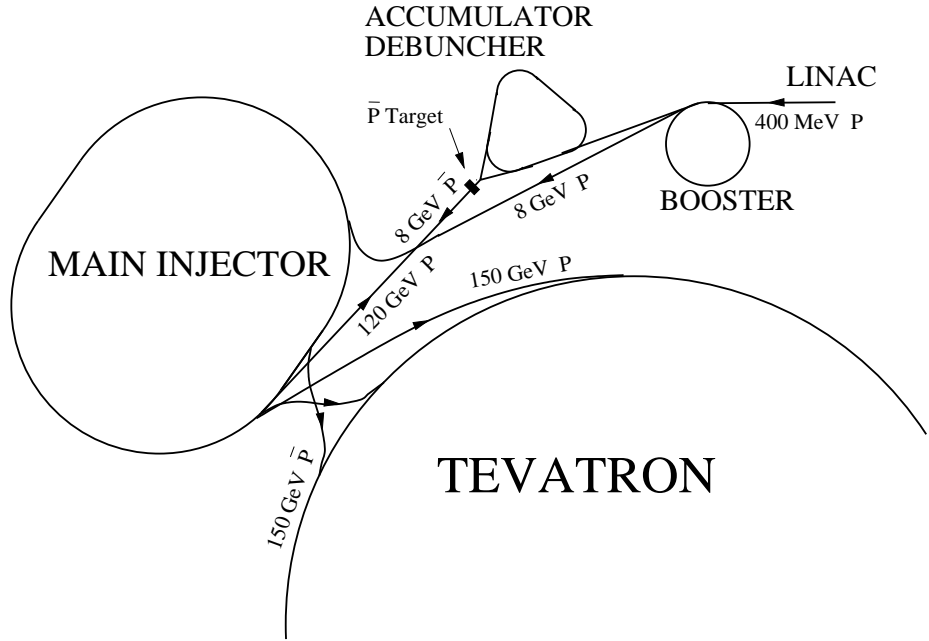}}
\end{picture}
\caption{Layout of the Fermilab accelerator complex for collider operations}
\label{fig:church}
\end{figure}

When all techniques alluded to in Sections~\ref{sec:twoexpt} to~\ref{sec:other} are
implemented, the integrated luminosity has still to be increased by a factor of 150
with respect to Run\,1 to reach a 5$\sigma$ sensitivity for $\mH\ = 115\,\Gcs$, 
and to extend the 95\% C.L. sensitivity domain beyond that already excluded by  
LEP\,2. Because the number of antiprotons drives the luminosity of a \ppbar\ collider,
it is necessary to produce, collect, handle and recycle many more antiprotons than at
Run\,1. These requirements imply a series of ambitious upgrades of the booster, 
the accumulator, the main injector, the transfer lines and 
the Tevatron itself 
(Fig.~\ref{fig:church}), some of which have already been completed, some of 
which are currently being commissioned, and some of which still entail large 
technical uncertainties. 

\noindent
A comprehensive description of these upgrades can be 
found in Ref.\cite{church}. Only a brief account of these 
improvements is given here.
\eject

\begin{itemize}
\item Booster upgrades for $\bar{\rm p}$ production

At Run\,1, the Booster proton intensity was limited by the integrated radiation 
losses, mostly due to beam losses. To reduce these losses, the Booster is being 
upgraded by {\it (i)} increasing the extraction aperture; {\it (ii)} reinforcing the
radiation shielding; and {\it (iii)} installing more corrector magnets to
improve the optics, and more beam collimators to localize losses in the safest 
areas.

\item Accumulation upgrades for $\bar{\rm p}$ collection

Once more protons are available from the booster, the next goal is to 
increase the number of antiprotons collected per proton on target. To 
do so, the Lithium-lenses magnetic focusing after the target was increased (by
increasing their gradient from 700 to 900\,T/m) and so was the accumulator 
aperture by beam pipe modification, larger septum magnet aperture, and improved 
beamline optics.

\item Cooling upgrades for $\bar{\rm p}$ handling

More antiprotons have then to be cooled, possibly faster than at Run\,1. The 
stochastic cooling system was therefore replaced by a brand new system with 
twice as large a bandwidth, which increases the cooling rate by a factor of 4. In 
addition, operating the cooling pickups at 4\,K drastically improves their 
signal-to-noise ratio and, therefore, their efficiency. Finally, electron cooling 
is foreseen in the main injector to deal with large antiproton intensities, 
for which stochastic cooling becomes less efficient.

\item Main injector upgrades for $\bar{\rm p}$ accumulation

Because the intensity of the proton source (prior to the target) is limited by 
space charge effects, two batches with small momentum offset are planned to be 
steered with two independent RF systems (with two different frequencies). The two 
resulting antiproton batches are merged later on in the main injector by 
bringing the two frequencies close to each other, thus allowing twice as large
a $\bar{\rm p}$ accumulation. Finally, it is planned to decelerate unused 
antiprotons in the Tevatron at the end of each collider store, so as to 
re-inject, keep and cool them in the Recycler for later use.
\end{itemize}

\noindent
With all the above  upgrades, the $\bar{\rm p}$ production rate and the number 
of antiprotons accumulated for each store  are expected to increase 
as shown in Table~\ref{tab:prodaccu}.

\begin{table}[htbp]
\caption{Antiproton production rates and numbers of $\bar{\rm p}$ produced 
for each collider store measured in Run\,1b and expected in Run\,2.}
\label{tab:prodaccu}
\begin{center}
\begin{tabular}{|l|c|c|c|}
\hline\hline
Run & Run\,1b (maximum) & Run\,2a (average) & Run\,2b (average) \\ 
Year & 1993-1996         & 2001-2003         & 2005-2007         \\ \hline\hline
Rate & $6 \times 10^{10}$ / hr &
$10 \times 10^{10}$ / hr &
$52 \times 10^{10}$ / hr \\ \hline
Number & $0.33 \times 10^{12}$ &
$1.1 \times 10^{12}$  &
$11 \times 10^{12}$  \\ \hline\hline 
\end{tabular}
\end{center}
\end{table}

The integrated luminosity expected from these upgrades is indicated in  
Table~\ref{tab:lumi}. With the proton and antiproton intensity increase, 
more bunches are needed to reduce the beam-beam tune shift, the background 
in the detectors and the number of interactions per crossing. Ultimately, 
the bunches will have to cross with a nonzero angle to keep the multiple 
interactions in the detectors to a manageable level, although it will 
reduce slightly the luminosity as well.

\begin{table}[htbp]
\caption{Integrated luminosity delivered per experiment (per week and total) 
in Run\,1b and expected to be produced per experiment in Run\,2.}
\label{tab:lumi}
\begin{center}
\begin{tabular}{|l|c|c|c|}
\hline\hline
Run & Run\,1b (maximum) & Run\,2a (average) & Run\,2b (average) \\ 
Duration & For one year  & For two years        & For three years  \\ \hline\hline
Config. & $6 \times 6$ bunches &
$36 \times 36$ bunches &
$140 \times 103$ bunches \\ \hline
Per week & 3.2\,\inpb & 17\,\inpb & 105\,\inpb \\ \hline\hline 
Total    & 0.14\,\infb & 1.5\,\infb & 14\,\infb \\ \hline\hline 
\end{tabular}
\end{center}
\end{table}

\subsection{Perspectives and Outlook}

The observability of a 115\,\Gcs\ Higgs boson at Tevatron Run\,2 relies on the 
realism of the performance ascribed to the foreseen improvements. 

On the 
one hand, some of the assumptions may look slightly optimistic: 
a 10\% dijet mass resolution was assumed, to be compared with 15\% measured 
in Run\,1, and 12\% expected in ATLAS and CMS; the aggressive assumptions on 
the b-tagging, neural network and trigger performance remain to  be demonstrated; 
a fast detector simulation was used throughout, although it is known to always 
give too good results; negligible systematic uncertainties were assumed, while any 
5\% systematic effect on the background would limit possible signal effects to 
2$\sigma$ for a typical signal-to-noise ratio of 10\%; the silicon trackers 
will have to be replaced in 2004, which requires a shutdown of the accelerator; 
and the integrated luminosity to be {\it collected} by CDF and D0 by 2007 
was assumed to 15\,\infb, which relies on the success of a solid, but very 
ambitious upgrading programme.

On the other hand, some of the assumptions are rather conservative: the analyses
used throughout are first-pass analyses, and may be improved; other relevant 
channels may contribute to Higgs production (e.g., \ttbar\H); the expected 
signal significance was computed with simple event counting, while events can
certainly be weighted ``\`a la LEP'' to improve the sensitivity; and LHC 
might even be further delayed, which would extend the  period during which 
15\,\infb\ have to  be accumulated by CDF and D0.

Although only future will tell us whether Run\,2 will be in a position
to confirm or not LEP's hints at 115\,\Gcs\ before the LHC, the present 
conjuncture is undoubtedly favourable to the Tevatron.

\section{Conclusion}

After twelve years of outstanding Physics at LEP, the precision electroweak 
measurements led to the prediction of the Higgs boson mass in the framework
of the standard model,
\begin{equation}
  \mH = 118^{+63}_{-42}\,\Gcs.
  \label{mHEW}
\end{equation}
More LEP running at high energy and at the Z pole would have allowed to reduce the 
uncertainty on the prediction from electroweak measurements to $\pm 15$\,\Gcs, which 
shows that LEP was stopped well before its Physics programme was over (as SLC was).
Direct searches for the HZ process unveiled an excess of signal-like events
corresponding to an almost 3$\sigma$ effect, compatible in every aspects with the 
production of a standard model Higgs boson of mass
\begin{equation}
  \mH = 115.0^{+0.7}_{-0.3}\,\Gcs,
  \label{mHDirect1}
\end{equation}
in remarkable agreement with Eq.~\ref{mHEW}. (After the conference, further analysis 
and systematic studies of the existing data, although still preliminary, confirmed 
qualitatively this effect at $115.6 \pm 1$\,\Gcs\ with a slightly reduced 
significance.) Six more months of LEP running in 2001 could have confirmed the 
hints and turn them into a 5$\sigma$ discovery. 

Instead, about five to ten years are now needed for a possible confirmation. Who is 
going to confirm is not yet clear: Many upgrades are still to be carried out to reach 
15\,\infb\ in 2007 at the Tevatron, and many costs are still to be covered to see
LHC starting in 2007.\cite{LHCWeb} The end of the decade may be thrilling.

\end{document}

%% file: commandes.tex
\newcommand{\btab}{\begin{tabular}}
\newcommand{\etab}{\end{tabular}}
\newcommand{\beq}{\begin{equation}}
\newcommand{\eeq}{\end{equation}}
\newcommand{\beqn}{\begin{eqnarray}}
\newcommand{\eeqn}{\end{eqnarray}}
\newcommand{\be}{\begin{equation}}
\newcommand{\ee}{\end{equation}}
\newcommand{\bea}{\begin{array}}
\newcommand{\ena}{\end{array}}
\newcommand{\begitm}{\begin{itemize}}
\newcommand{\eitm}{\end{itemize}}

\def\epem{\mbox{$e^+e^-$}}

\def\Gcs{\mbox{GeV/$c^2$}}

\def\infb{\mbox{$\mbox{fb}^{-1}$}}
\def\inpb{\mbox{$\mbox{pb}^{-1}$}}

\def\epem{\mbox{$e^+e^-$}}

\newcommand{\AmS}{{\protect\the\textfont2
  A\kern-.1667em\lower.5ex\hbox{M}\kern-.125emS}}

\def\inpb{\mbox{$\hbox{pb}^{-1}$}}
\def\infb{\mbox{$\hbox{fb}^{-1}$}}

\def\Gcs{\hbox{GeV}/\mbox{$c^2$}}

\def\Z{\mbox{$\hbox{Z}$}}

\def\W{\mbox{$\hbox{W}$}}

\def\H{\mbox{$\hbox{H}$}}

\def\WpWm{\mbox{$\W^+\W^-$}}

\def\g{\mbox{$\hbox{g}$}}
\def\epemto{\mbox{$\hbox{e}^+\hbox{e}^- \to$}}
\def\epem{\mbox{$\hbox{e}^+\hbox{e}^-$}}
\def\mpmm{\mbox{$\mu^{+}\mu^{-}$}}
\def\tptm{\mbox{$\tau^+\tau^-$}}
\def\lplm{\mbox{$\ell^+\ell^-$}}
\def\nnbar{\mbox{$\nu\bar\nu$}}
\def\ffbar{\mbox{$\hbox{f}\bar{\hbox{f}}$}}

\def\qqbar{\mbox{$\hbox{q}\bar{\hbox{q}}$}}
\def\qqpbar{\mbox{$\hbox{q}\bar{\hbox{q}}^\prime$}}

\def\ppbar{\mbox{$\hbox{p}\bar{\hbox{p}}$}}
\def\bbbar{\mbox{$\hbox{b}\bar{\hbox{b}}$}}
\def\ttbar{\mbox{$\hbox{t}\bar{\hbox{t}}$}}

\def\mH{\mbox{$m_{\hbox{\eightrm H}}$}}

\def\mZ{\mbox{$m_{\hbox{\eightrm Z}}$}}

\def\suppeg{\hbox{\lower -.08cm \hbox{
{\hbox{$>$}}{\hbox{\kern -.30cm\lower .18cm \hbox{$\sim$}}}}}}
\def\infpeg{\hbox{\lower -.08cm \hbox{
{\hbox{$<$}}{\hbox{\kern -.50cm\lower .28cm \hbox{$\sim$}}}}}}
